\newcommand{\ket}[1]{|#1\rangle}
\newcommand{\up}{\uparrow}
\newcommand{\down}{\downarrow}
\renewcommand{\H}{\cal H}
\newcommand{\G}{\cal G}
\newcommand{\F}{\cal F}
\newcommand{\be}{\begin{equation}}
\newcommand{\ee}{\end{equation}}
\newcommand{\bea}{\begin{eqnarray}}
\newcommand{\eea}{\end{eqnarray}}
\documentstyle[aps,prl,multicol]{revtex}
\begin{document}
\title{An analytically solvable time dependent Jaynes~Cummings~Model}
\author{Ananda Dasgupta}
\address{Saha Institute of Nuclear Physics,1/Af Bidhannagar,Calcutta - 700064,
	India}
\maketitle
\begin{abstract}
Using the underlying su(2) algebra of the Jaynes-Cummings Model (JCM), we 
construct a time dependent interaction term that allows analytical 
solution for even off-resonance conditions. Exact solutions for the time 
evolution of any state has been found. The effect of detuning on the Rabi
oscillations and the collapse and revival of inversion is indicated. 
It is also shown that at resonance, the time dependent JCM is analytically 
solvable for an arbitrary interaction term.
\end{abstract}
\section{Introduction}
The Jaynes Cummings Model (JCM) \cite{JC,KR} is one of the major paradigms in 
quantum optics today. It was introduced as a highly simplified model to explain, 
qualitatively, the salient features of the interaction of matter with quantised 
radiation field in a cavity. Despite the fact that it considers matter in the 
highly simplified form of a two level atom and radiation in the form of a single 
mode field it demonstrates quite a few uniqely quantum mechanical features. It's 
importance has been enhanced recently by experimental realisation in Rydberg 
atom masers \cite{MWG,RW}. Various
generalisations of the basic JCM have been considered in the literature - 
multiphoton interactions, nonlinear media, deformed radiation fields to name 
but a few. Recently, a few authors have treated the time dependent
version of the JCM \cite{PY,JL1,JL2,GAC,GACP1,GACP2}. In this paper we treat a 
particular class of time dependent JCM's, where the time dependence of the 
interaction term is such that it allows for an analytical solution.
	
This article is organised as follows. In section II, we discuss the 
generalisation of the basic JCM due to Yu {\em et al}, \cite{YRZ}, which 
encompasses most of the generalized JCMs that have been treated in the 
literature. In section III we discuss, following \cite{YRZ}, the underlying 
su(2) algebra that allows a reduction of the generalized JCM into the problem
of the dynamics of a spin-$\frac{1}{2}$ particle in amagnetic field. In this
article, we are concerned with time dependent interaction, which will lead to a
generalisation over \cite{YRZ} in that it leads to a time dependent magnetic 
field. Here we point out why the case of zero detuning, which has been treated
by most authors in the context of the time dependent model, is analytically 
solvable for all choices of the time dependence.
In section IV, we describe our choice for the time dependence of the
interaction term, for which an analytic solution can be obtained for arbitrary 
detuning . We also outline a possible physical situation where such an
interaction could arise. Section V contains details of the analytic solution, 
while section VI discusses the time evolution of atomic inversion in the 
one-photon linear medium case, as an example of the results that can be 
obtained. Finally, section VII has a few concluding remarks, and a few avenues
for further investigations are indicated.

\section{The generalised JCM }
The basic JCM is defined by the Hamiltonian
\begin{equation}
H=\omega a^\dagger a + \frac{\omega_0}{2} \sigma_3 
+ \left(\lambda a^\dagger\sigma_- +\lambda^*a\sigma_+\right) 
\end{equation}
where $a$ is the annihilation operetor for the single mode radiation field,
while $\sigma_3,\sigma_-,\sigma_+$ are pseudospin operators describing the two
level atom. Various authors have considered different generalisations to this 
basic Hamiltonian. 
In what follows we are going to consider the case of real coupling
constant $\lambda(t)=\lambda(t)^*$.

The wide variety of generalized JCM Hamiltonians that can be exactly solved, at
least for time independent couplings, have led Yu {\em et al} (\cite{YRZ})into 
a search for an underlying algebraic structure unifying all of them. In their 
notation a very general class of JCMs can be represented by the generic formula 
\begin{equation}
H= r\left(A_0\right)+s\left(A_0\right)\sigma_3+\lambda(t)\left(A_+\sigma_-+A_-\sigma_+\right)
\label{ham}
\end{equation}
where $A_0$ is some generalised number operator for the photon field, and $A_+,
(A_-)$ represent generalised raising (lowering) operators for it. We have 
changed the original notation slightly in order to exhibit the time dependence 
in the interaction term explicitly. Here the 
first term in the Hamiltonian (\ref{ham}) obviously stands for the energy of a 
free photon field, whereas the second term stands for the two level atom, the
level splitting depending on the strength of the photon field. The last two 
terms stand for the field-atom interaction in the rotating wave approximation.
Here, the generalisation over the standard JCM is that the operators $A_+,(A_-)
$ raise (lower) the eigenvalue of $A_0$ by a number m which can be integer or 
fractional. Obviously a rather large collection of generalized JCMs, including
those mentioned above, can be treated as special cases of this Hamiltonian.

As noted in \cite{YRZ} the algebraic structure of the system can be summarised
by 
\begin{eqnarray}
[A_0,A_\pm] &= \pm m A_\pm\\
A_+A_- &= \chi(A_0)
\end{eqnarray} 
By following the same elementary steps as for determination 
of the angular momenta eigenvalues from their algebra, here we can see that the
states for which $\chi(A_0)$ vanishes, while increasing, which we will 
collectively refer to as "low'' states act as lower cut off for ladders of 
states for which  the $A_0$ eigenvalues change in steps of m. Each state of the
ladder is generated from the previous one by the action of $A_+$. Depending on 
the function $\chi(n)$, the ladders may or may not have a upper cut off. We 
will refer to the upper cut off states, if they exist as "high'' states.
 
\section{The underlying algebra}
One next notes that the Hamiltonian (\ref{ham}) admits of a conserved quantity
\begin{equation}
 \Delta = A_0 +\frac {\left( 1+\sigma_3\right)}{2}m
\end{equation}
which not only commutes with the Hamiltonian, does so with every one of it's 
four terms. Thus the dynamics splits into a number of
two dimensional subspaces spanned by $\ket{n,\up}$ and $\ket{n+m,\down}$, the
first label being the $A_0$ eigenvalue. On the other hand, the states
$\ket{l,\down}$ and $\ket{h,\up}$ where the $A_0$ eigenstates $\ket{l}$ and
$\ket{h}$ belong to the "low'' and "high'' states, respectively span one 
dimensional invariant subspaces under the dynamics. Time evolution of these 
latter subspaces is trivial, since the interaction term vanishes there.

Leaving aside the one-dimensional subspaces, where the operator $\chi(\Delta)$
vanishes, we can define three spin operators 
\begin{eqnarray}
J_1 =& \frac{1}{2\sqrt{\chi(\Delta)}}\left(A_+\sigma_-+A_-\sigma_+\right)\\
J_2 =& \frac{i}{2\sqrt{\chi(\Delta)}}\left(A_+\sigma_--A_-\sigma_+\right)\\
J_3 =& \frac{1}{2}\sigma_3
\end{eqnarray}
These operators obey the pauli matrix algebra
\begin{equation}
 J_iJ_j=i \epsilon_{ijk} \frac{J_k}{2}+\frac{\delta_{ij}}{4}I.
\end{equation}
Now, one can rewrite (\ref{ham}) in terms of $\Delta$ :
\be
H= \Omega(\Delta) + \delta(\Delta)\sigma_3 +\lambda(t)\left(A_+\sigma_-
+A_-\sigma_+\right)
\ee
Here the terms in this Hamiltonian is related to those in the original one by
\bea
\Omega(\Delta)=&\frac{r(\Delta-m)+r(\Delta)}{2}+\frac{s(\Delta-m)-s(\Delta)}{2}\\
\delta(\Delta)=&\frac{r(\Delta-m)-r(\Delta)}{2}+\frac{s(\Delta-m)+s(\Delta)}{2}
\eea

Before proceeding further, we give a specific example for the sake of 
concreteness. We consider the case of the m-photon JCM in a Kerr medium. This is 
described by (\ref{ham}) with
\bea
A_0 =& a^\dagger a\nonumber\\
A_+ =& (a^\dagger)^m\nonumber\\
A_- =& a^m\nonumber\\
\chi(n)=& \frac{n!}{(n-m)!}\nonumber\\
r(n)=& \omega n +\kappa(n^2-n)\nonumber\\
s(n)=& \frac{\omega_0}{2}\nonumber
\eea
This leads to
\bea
\Omega(\Delta) =& \left(\Delta -\frac{m}{2}\right)\omega
	+\kappa\left(\Delta^2-(m+1)\Delta+\frac{m(m+1)}{2}\right)\nonumber\\
\delta(\Delta)=&\frac{\delta}{2}
	+\kappa\left(\frac{m(m+1)}{2}- m\Delta\right)\nonumber
\eea
where the detuning parameter $\delta$ is defined as $\omega_0 - m\omega$. In 
this example the photon number states $\ket{0},\ket{1},...\ket{m-1}$ comprise
the low states $\ket{l}$, and there are no high states. 
Putting $\kappa=0$ gives us the m-photon JCM.

Resuming our analysis, the time evolution operator can be written as
\be
U(t)= e^{-i\Omega(\Delta)(t-t_0)}U_i(t)
\ee
with $U_i$ satisfying the interaction picture Schr\"{o}dinger equation
\be
i\frac{\partial U_i}{\partial t} = H_i(t) U_i(t)
\ee
where, in terms of the pseudospin operators the interaction Hamiltonian becomes
that of a negatively charged spin-$\frac{1}{2}$ in a time varying magnetic
field:
\begin{equation}
 H_i={\bf J\cdot B},
\end{equation}
where the magnetic field {\bf B} is given by:
\begin{equation}
{\bf B} = 2\left(\lambda(t)\sqrt{\chi(\Delta)},0,\delta(\Delta)\right)
\end{equation}
In the time independent JCMs, this reduces to
the motion of a spin-$\frac{1}{2}$ particle in a constant magnetic field
the case that was treated in \cite{YRZ}.
Our interest lies in the case where the coupling is time dependent. It is easy 
to see that for JCMs in linear media the case of zero detuning $\delta=0$ is 
special in that there the magnetic field is uni-directional. Thus the 
interaction Hamiltonian at different times commute, leading to a exactly
solvable time evolution operator.
\be
U_i=\exp\left[2i\sqrt{\chi(\Delta)}\left(\int_{t_0}^tdt'\  \lambda(t')\right)
	J_1\right].
\ee
Most of the authors who have treated time dependent versions of the JCM 
analytically have dealt with the case of zero detuning \cite{PY,JL1,JL2}. It
should be noted, however, that in the case of presence of Kerr or other
nonlinearities, the function $\delta(\Delta)$ would not reduce to a constant, 
and the magnetic field will not be unidirectional for all the invariant 
subspaces even for zero detuning. 
  
	Our interest in the current article lies in looking for a time 
dependent interaction term, which will allow analytical solution even when the
detuning is non-zero. In this we benefit from the fact that the spin - field
interaction has been widely studied.

\section{The time dependence}
In our version of the JCM, we take the coupling constant $\lambda$
to be time dependent of the form :
\begin{equation}
\lambda(t) = \lambda_0 {\rm sech}(t/(2\tau)).\label{int}
\end{equation}
In this form the coupling increases from a very small value at large negative
times to a peak at time $t=0$, to decrease exponentially at large times. Thus,
depending on the value of $\tau$ and the initial time $t_0$, various limits 
such as adiabatically or rapidly increasing (for $t_0<t<=0$) or decresing 
(for $0<=t_0<t$) coupling can be conveniently studied. This allows us to 
investigate, analytically, the effect of transients in various different
limits of the effect of switching the interaction on and off in the cavity
atom-maser system. 
 The choice of the interaction was guided by the fact that the
JCM has su(2) as an underlying algebra \cite{YRZ}, and the above interaction is 
known to give rise  to exactly solvable dynamics for a spinning particle in a
time dependent magnetic field \cite{DGT,RZ}.

It should be noted that the time dependence specified in (\ref{int}) is only 
one of a class of generalised interactions that afford analytical solutions
\cite{BB}.
One reason why we confine ourselves to this particular interaction is that it 
is the simplest one among this class. Also, the form of the 
potential (\ref{int}) is not unfamiliar in the subject of two level atoms 
interacting with radiation. The famous area and shape stable solution 
for the loss less Maxwell equation coupled with the optical Bloch equation that
is encountered in the study of self induced transparency has the same time 
dependence \cite{MH,AE}. Phase modulation effects in such a medium  is another 
place where this shows up \cite{MH,AE}. One possible use of this potential, 
then, could be the investigation of  the effects of quantizing the radiation 
field in these problems.

\section{The analytical solution}
Following the Wei-Norman formalism \cite{DGT,WN} we write the evolution operator 
in the form (where we have suppressed the $\Delta$ dependence for convenience) 
\begin{equation}
U_i=e^{-i\delta(t-t_0)\sigma_3}\tilde{U_i},
\end{equation}
where $\tilde{U_i}$ is 
\begin{equation}
\tilde{U_i}=e^{h(t)\sigma_3}e^{g(t)J_+}e^{-f(t)J_-}.
\end{equation}
We introduce the new functions
\begin{eqnarray}
\H&=e^{-h},\nonumber\\
\F&=f\ e^{-h},\\
\G&=g\ e^h.\nonumber
\end{eqnarray}
It can be shown that $\H,\G^*$ and $\F$ obeys the second order differential 
equation
\be
\frac{d^2X}{dt^2} + \left[-\frac{d}{dt}\ln\lambda(t) +i2\delta(n+m)\right]
	\frac{dX}{dt} +\chi(n+m)\lambda(t)^2X=0,\label{deq}
\ee
with the initial conditions
\bea
{\H}(t_0)=&1,\nonumber\\
\dot{\H}(t_0)=&0,\nonumber\\
{\F}(t_0)=&0,\nonumber\\
\dot{\F}(t_0)=&i\sqrt{\chi(n+m)}\lambda(t_0).\nonumber
\eea
The initial conditions for $\F$ and $\G^*$ are identical, implying
\be
\G=\F^*.
\ee
In terms of these new functions the operator $\tilde{U_i}$ is
\begin{equation}
\tilde{U_i}=\left(\begin{array}{cc}
			\H^*&\F^*\\
			-\F&\H\end{array}\right).
\end{equation}

Changing variable to
\be
z(t)=\frac{e^{t/\tau}}{1+e^{t/\tau}},
\ee
the equation~(\ref{deq}) becomes
\be
z(1-z)\frac{d^2X}{dz^2}+(\gamma-z)\frac{dX}{dz}+\alpha^2 X=0,
\label{deq2}
\ee
where $\gamma=1/2+i2 \delta(n+m)\tau$ and $\alpha=2\lambda_0\tau\,\sqrt{\chi(n+m)}$.
This is the hypergeometric equation with $\beta=-\alpha$ and
thus 
\be
{\cal H}=A_h\ _2F_1(\alpha,-\alpha;\gamma;z) + B_h\ z^{1-\gamma}
	\ _2F_1(\alpha-\gamma+1, -\alpha-\gamma+1;2-\gamma;z)
\ee
\be
{\cal F}=A_f\ \ _2F_1(\alpha,-\alpha;\gamma;z) + B_f\ z^{1-\gamma}
	\ _2F_1(\alpha-\gamma+1, -\alpha-\gamma+1;2-\gamma;z)
\ee
The initial conditions lead to~:
\bea
A_h =&\left(1-z_0\right)^{1-\gamma}
	 \ _2F_1(\alpha-\gamma+1, -\alpha-\gamma+1;1-\gamma;z_0),\nonumber\\
B_h =&\frac{\alpha^2 z_0}{\gamma(1-\gamma)}\left(\frac{z_0}{1-z_0}
	\right)^{\gamma-1} \ _2F_1(\alpha+1,-\alpha+1;\gamma+1;z_0),\nonumber\\
A_f =&-\frac{i\alpha}{1-\gamma}\left(\frac{z_0}{1-z_0}\right)^{\gamma-1/2}
	z_0^{1-\gamma}\ _2F_1(\alpha-\gamma+1,-\alpha-\gamma+1;2-\gamma;z_0),
	\nonumber\\
B_f =&\frac{i\alpha}{1-\gamma}\left(\frac{z_0}{1-z_0}\right)^{\gamma-1/2}
	\ _2F_1(\alpha,-\alpha;\gamma;z_0)
\eea
where $z_0=z(t_0)$. 

If we now write the state of the system at time t, by
\be
\ket{\psi(t)}=\sum_{n=0}^{\infty}\left[u_n(t)\ket{n,\up}+v_n(t)\ket{n,\down}
	\right],
\ee
the time dependent functions $u_n(t)$ and $v_n(t)$ can be seen to be
\bea
u_n(t)=&\exp\left[- i\left(\Omega(n+m)+\delta(n+m)\right)(t-t_0)
	\right]\ \ \left[{\cal H}^*_{(n+m)}u_n(t_0)+{\cal F}^*_{(n+m)}
	v_{n+m}(t_0)\right],\nonumber\\	
v_{n+m}(t)=&\exp\left[- i\left(\Omega(n+m)-\delta(n+m)
	\right)(t-t_0)\right]\ \ \left[-{\cal F}_{(n+m)}u_n(t_0)+
	{\cal H}_{(n+m)}v_{n+m}(t_0)\right].\label{evol}	
\eea
Here the subscripts on $\H$ and $\F$ denote that they correspond to the subspace
$\Delta=n+m$,{\em i.e.} the subspace spanned by $\ket{n,\up}$ and $\ket{n+m,
\down}$. The evolution equation~(\ref{evol}) is valid for all nonnegative
integer $n$. The one dimensional subspaces, spanned by the "low'' states, 
$\ket{l,\down}$ evolve according to
\be
v_l(t)=e^{-i(\Omega(l)-\delta(l))(t-t_0)}\ v_l(t_0)\label{evol'}.
\ee
while the evolution of the "high'' states $\ket{h,\up}$ is given by
\be
u_h(t)=e^{-i(\Omega(h+m)+\delta(h+m))(t-t_0)}\ u_h(t_0)\label{evol''}.
\ee
The equations (\ref{evol}),(\ref{evol'}) and (\ref{evol''}) describe the 
complete time evolution of the state of the Jaynes Cummings system.

\section{Time evolution of atomic inversion}
Although all the results presented so far are equally applicable to all the
generalized JCM's that  can be described by (\ref{ham}) in what follows we
will deal only with the standard one-photon JCM with time dependent coupling 
given by (\ref{int}) mainly for the sake of brevity.
A quantiy of physical interest is  the time evolution of the atomic inversion 
$\sigma_3$ . This is given, in the one-photon case, by,
\be
\langle\sigma_3(t)\rangle=-|v_0(t_0)|^2+\sum_{n=0}^{\infty}\left[
	\left(1-2|{\cal F}_{n+1}|^2\right)
	\left(|u_n(t_0)|^2-|v_{n+1}(t_0)|^2\right) +4{\rm Re}\left(
	{\cal H}_{(n+1)}{\cal F}^*_{(n+1)}u_n(t_0)^*v_{n+1}(t_0)\right)\right],
\ee 
where use has been made of $|{\cal H}_{n+1}|^2+|{\cal F}_{n+1}^2|=1$.
In particular, we consider the initial condition where the radiation field is 
in the number state $\ket{n}$ and the atom is in a superposition 
$c_e\ket{\up}+c_g\ket{\down}$, the atomic inversion evolves as
\be
\langle\sigma_3(t)\rangle=p_e\left(1-2|{\cal F}_{(n+1)}|^2
	\right) -(1-p_e)\left(1-2|{\cal F}_{(n)}|^2\right).
\ee
A uniquely quantum mechanical feature of the Jaynes-Cummings model is the
collapse and revival of atomic inversion when the radiation field is initially 
in a coherent state. We consider a special case when the atom is initially in 
the excited state, and derive
\be
\langle\sigma_3(t)\rangle=\sum_{n=0}^\infty\left(1-2|{\cal F}_{(n+1)}|^2\right)
	\frac{\bar{n}^n}{n!}e^{-\bar{n}},
\ee
where $\bar{n}$ is the mean photon number in the coherent state.

In Fig. 1 we plot the variation of the atomic inversion with time, for 
initial time $t_0=-10\tau$,for two different detunings $\delta=0$ and $\delta
=\tau$, respectively. In this, we have 
assumed the initial condition that the atom is in the excited state, $p_e=1$,
and the radiation field is in the number state $n=3$. The effect of the detuning
on both the frequency and the amplitude of the Rabi oscillations is quite 
marked. Note, that as the coupling strenghth varies, the frequency of Rabi
oscillations (the Rabi frequency) changes. The vanishing of
the interaction at large positive times leads to the levelling out of the 
inversion. Fig. 2 shows the inversion for an inital state where the atom is 
excited and radiation is in a coherent state $\bar{n}=10$ for $\lambda_0\tau
=5.0$ for various detunigs. The initial time is $t_0=-10\tau$, so the 
interaction starts at a fairly low value, peaks and then drops off again.
The inversion shows a single revival and collapse after the initial collapse,
finally levelling out at large times as expected. Finally Fig. 3 shows the
same quantities, but with $t_0=0$, so that the interaction decreases 
monotonically. It is seen that the interaction drops off too fast to show
secondary collapses, after the initial rapid one. It is also seen that though
the effect of the detuning is quite marked for $t_0=-10\tau$ it is small for
$t_0=0$.

\section{Conclusions}
	We have seen one example of a time dependent JCM which can be solved 
analytically for arbitrary detunings. Of course, a large number of other
physical quantities, {\em e. g.} photon number distribution, squeezing of
the radiation field, atom-field relative phase can be studied for this
particular model. Of particular interest may be the effect of introducing
a Kerr nonlinearity in the problem.

As pointed out before, (\ref{int}) is one
member of a large subclass of interactions which have been studied in
the NMR literature which lead to solvable time dependent problems. In 
particular, one can consider cases where the time
dependence is assymetric about $t=0$, and investigate the effects of such
assymetry on physical observables.

	Apart from the fact that the reduction of the Jaynes Cummings system
to the spin - magnetic field system helps us to find interactions for which 
analytical solutions can be found, the algebraic structure also allows us to 
develop a convenient approximation scheme for general time dependences. We can 
treat the case of small detuning as a perturbation on the exactly solvable zero
detuning case. Algebraic unitarity preserving perturbation methods, such as the
Magnus-Fer method \cite{DGT,M} could be particularly convenient here. 
Such extensions will be the subject of a future article.

The routine developed by Perger {\it et al} \cite{PBN} was used to calculate
the hypergeometric function.

\newpage
\centerline{\bf \large{Figure Captions}}
\vskip 1 cm
\begin{itemize}
\item{\bf Figure 1 -} Atomic inversion for an initially excited atom, with the 
radiation field in a number state $n=3$, at initial time $t_0=-10\tau$.  
The peak strength of the interaction is $\lambda_0=5\tau$. The solid line
describes the system at resonance, $\delta=0$, while the broken line is for
$\delta= \tau$.
\item{\bf Figure 2} Atomic inversion for $\lambda_0=5\tau$ 
for an initially excited atom and the radiation field in a coherent state 
with $\bar{n}=10$ at the initial time $t_0 =-10\tau$. The solid line is for
$\delta=0$ and the broken line is for $\delta=0.5\tau$.
\item{\bf Figure 3} Same as in Fig. 2, except for $t_0=0$.
\end{itemize}
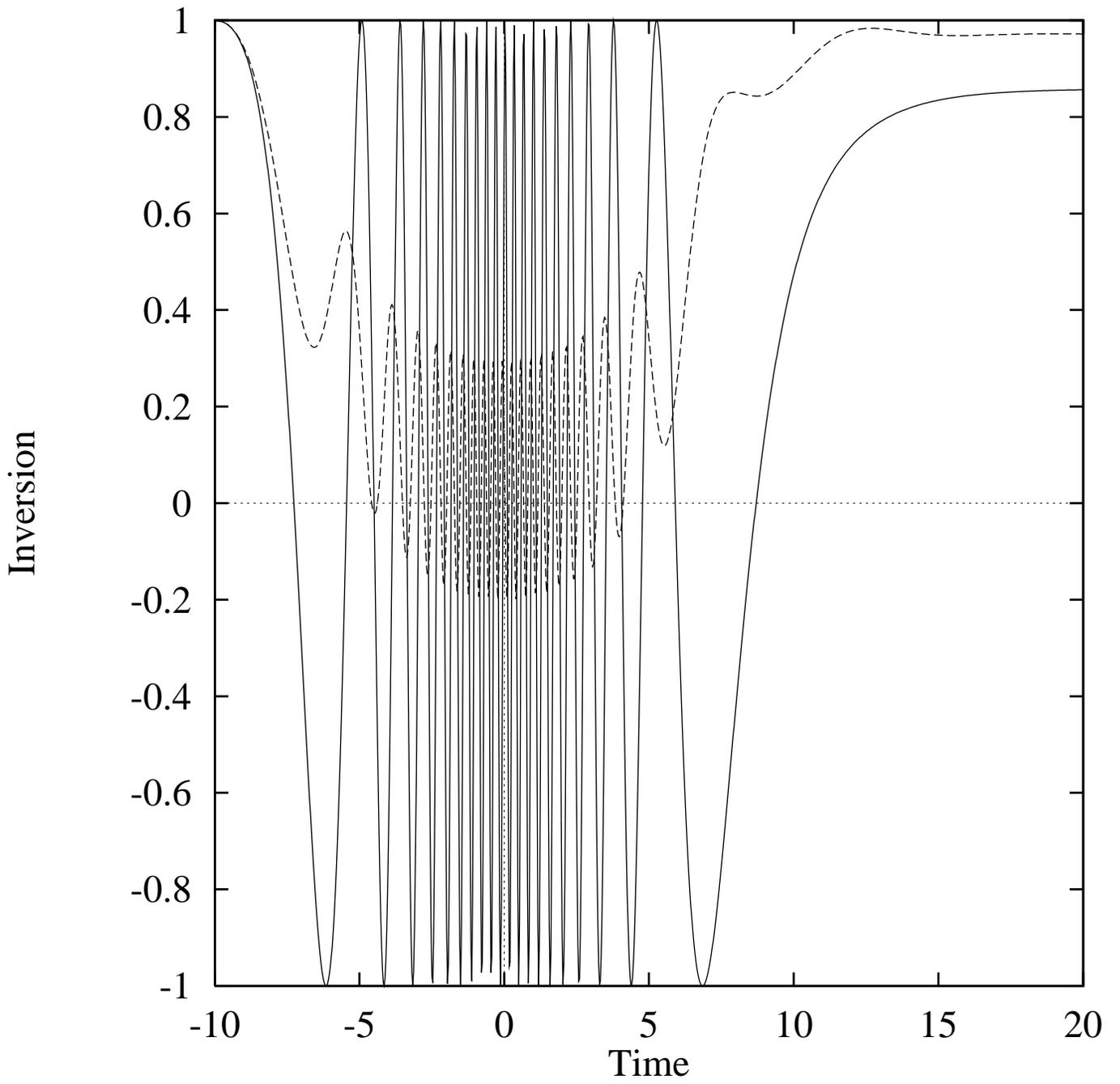
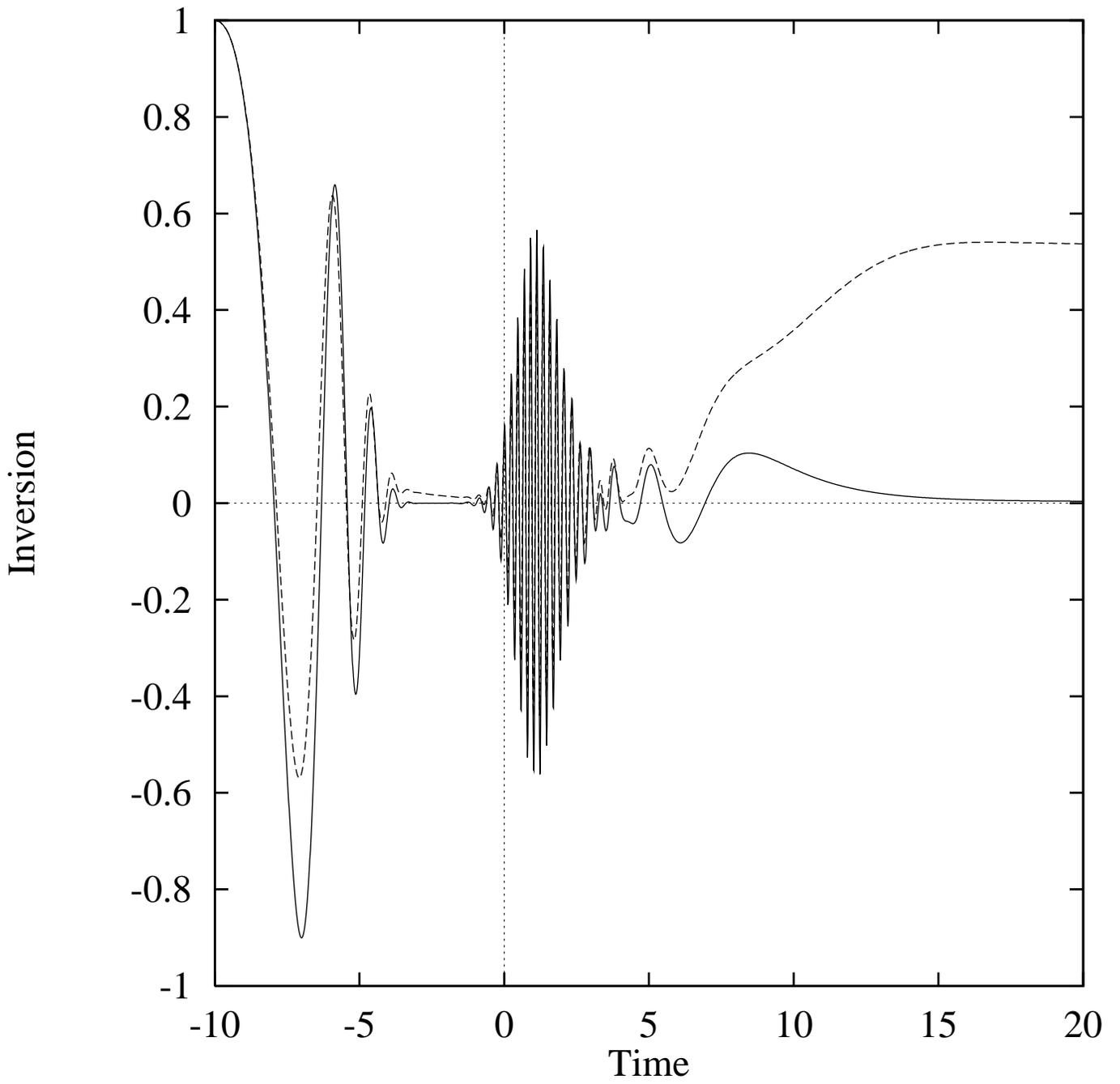
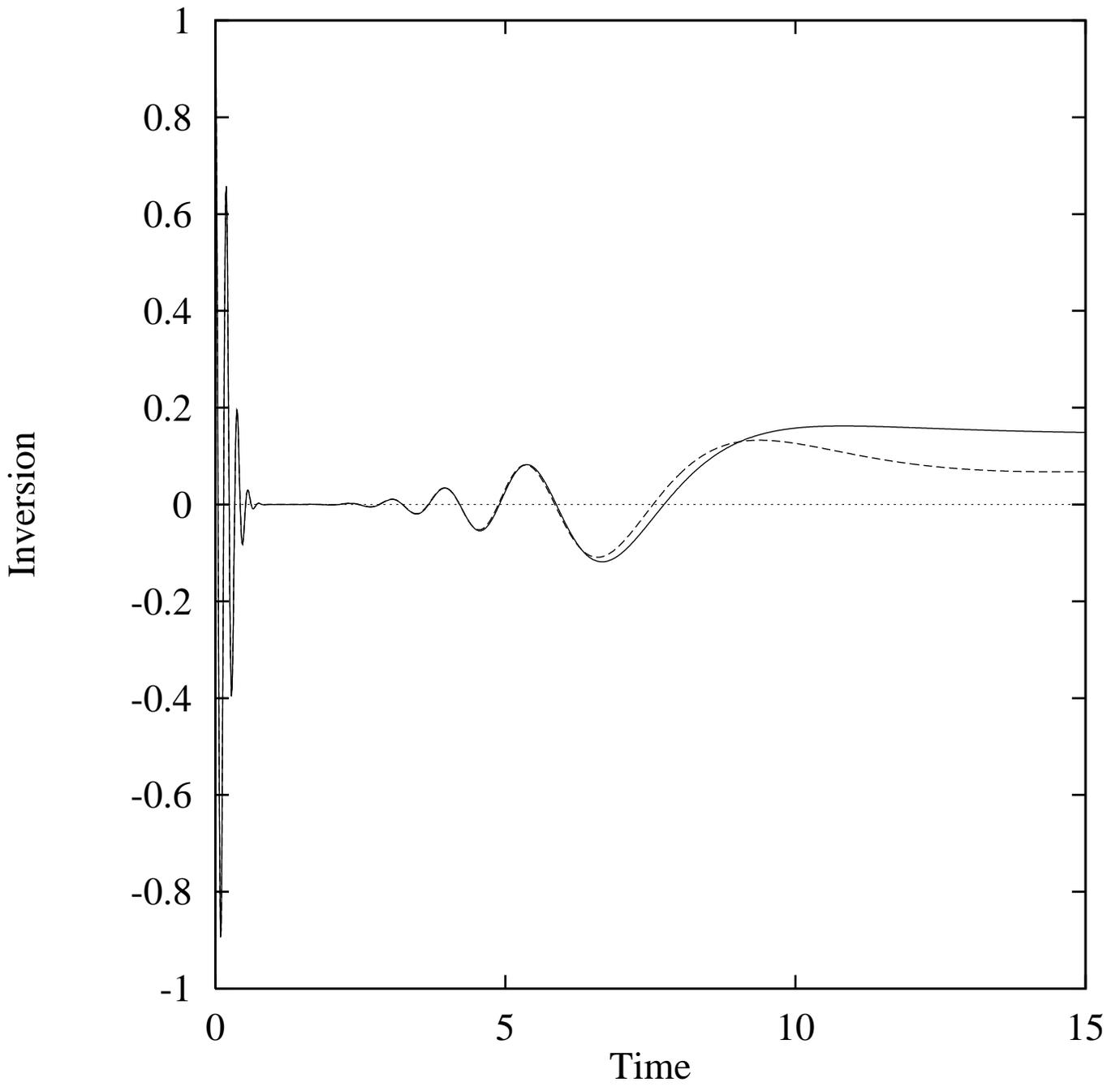
\end{document}